\documentclass{article}
\usepackage[utf8]{inputenc}
\usepackage[margin=20mm]{geometry}
\usepackage{amsmath,amssymb}
\usepackage{algorithm,algorithmic}
\usepackage{graphicx}
\usepackage{bm}
\usepackage{color}
\usepackage{comment}
\usepackage{subcaption}
\usepackage{authblk}
\usepackage{hyperref}

\newcommand{\wt}{\widetilde}
\newcommand{\changed}[1]{{\sf\color[rgb]{1,0,0}{#1}}}
\newcommand{\com}[1]{{\sf\color[rgb]{0,0,1}{#1}}}

\DeclareMathOperator*{\argmin}{arg\,min}

\title{Asymptotic evaluation of the information processing capacity in reservoir computing}
\author{Yohei Saito \thanks{\href{mailto:saito.yohei450@mail.kyutech.jp}{saito.yohei450@mail.kyutech.jp}}}
\affil{Graduate School of Life Science and Systems Engineering, Kyushu Institute of Technology}
\date{}

\begin{document}
\maketitle
\begin{abstract}
Reservoir computing (RC) is becoming increasingly important because of its short training time. 
The squared error normalized by the target output is called the information processing capacity (IPC) and is used to evaluate the performance of an RC system. 
Since RC aims to learn the relationship between input and output time series, we should evaluate the IPC for infinitely long data rather than the IPC for finite-length data. 
However, a method for estimating it has not been established. 
We evaluated the IPC for infinitely long data using the asymptotic expansion of the IPC and weighted least-squares fitting. 
Then, we showed the validity of our method by numerical simulations. 
This work makes the performance evaluation of RC more evident. 
\end{abstract}

\section{Introduction}
Since many kinds of data, e.g. video, audio, and stock prices, have time correlation, machine learning of time-series data is an important issue. 
Recurrent neural networks (RNNs) can store past input by recursively connecting hidden nodes~\cite{rumelhart1986learning} and can approximate the relationship between input and output time series with arbitrary accuracy~\cite{schafer2006recurrent}. 
Backpropagation through time (BPTT) is mainly used to train RNNs, but it is difficult to optimize network parameters due to the gradient vanishing or the gradient explosion~\cite{bengio1994learning}. 
Many variants of RNNs, such as LSTM~\cite{hochreiter1997long} and GRU~\cite{cho2014learning}, have been proposed to solve the difficulty of training and have been very successful. 
However, training with BPTT takes a long time. 

An echo state network (ESN)~\cite{jaeger2004harnessing} is a kind of RNNs, which can finish training quickly by fixing the recurrent connections at the initial value and optimizing only the linear transformation of the readout layer. 
Not limited to neural networks, a linear combination of nonlinear dynamical systems can be used to approximate the relationship between input and output time series and is called a reservoir computing (RC) system~\cite{tanaka2019recent}. 
RC systems can also approximate the relationship between input and output time series with arbitrary accuracy~\cite{grigoryeva2018echo, gonon2019reservoir}. 
Since no optimization is performed other than the linear transformation, an RC system is often inferior in performance to LSTM and other methods. 
However, RC has the advantage that training is finished quickly by simply calculating the pseudoinverse matrix. 
It has different demands than large-scale neural networks, which require large amounts of data and high computational costs. 
Recently, RC has been the subject of active research~\cite{yan2024emerging}, including improvements in reservoir dynamics~\cite{palacios2024role, peng2024reservoir}, physical implementation~\cite{gao2024toward, lee2024task}, and theoretical analysis~\cite{ballarin2025memory}. 



The performance of an RC system is evaluated by the mean squared error (MSE) or the squared error normalized by the target output, and the latter is called the information processing capacity (IPC)~\cite{dambre2012information, kubota2021unifying}. 
Through normalization, the IPC takes a value between $0$ and $1$, making it an intuitive index of approximation accuracy. 
Since the goal of RC is to learn the relationship between input and output time series, it is necessary to evaluate the IPC for infinitely long data, not the IPC for finite-length data. 
Previous studies have been careful not to overestimate the IPCs that vanish in the infinite time limit, but those that do not vanish in the infinite time limit are substituted with the IPCs for sufficiently long data~\cite{dambre2012information, kubota2021unifying, vettelschoss2021information, schulte2023refined, takasu2025neuronal}. 
However, the difference between the IPC for finite and infinite data lengths still exists and should be removed as much as possible. 
In this paper, we estimate the IPC for infinitely long data using the asymptotic expansion of the IPC and the weighted least squares method. 
We also propose a robust method to determine whether the IPC becomes zero in the infinite time limit. 

This paper is organized as follows. 
Section 2 reviews RC and its performance index, the IPC. 
In Section 3, we derive the asymptotic form of the IPC and explain the method for estimating the IPC. 
In Section 4, we show the validity of our method by numerical simulations. 
Section 5 summarizes the paper. 

\section{Review of RC and the IPC}  
The dynamics of RC is expressed as follows. 
\begin{align}
 x_t =& f(x_{t-1}, u_t),
 \label{eq:reservoir_nodes} \\
 y_t =& w^\top x_t + b,
 \label{eq:reservoir_output}
\end{align}
Here, $u_t \in \mathbb{R}^{d_0}, x_t \in \mathbb{R}^{d_1}, y_t \in \mathbb{R}^{d_2}$ are the values of the input, the hidden nodes, and the reservoir output at time $t$, respectively. 
To simplify the notation, we rewrite $(x_t^\top, 1)^\top$ as $x_t$, $(w^\top, b)^\top$ as $w$, and Eq.~(\ref{eq:reservoir_output}) can be rewritten as 
\begin{align}
 y_t =& w^\top x_t.
 \label{eq:reservoir_output_2}
\end{align}
To obtain $x_t$ from Eq.~(\ref{eq:reservoir_nodes}), the values of the hidden nodes at a certain time $-\tau \, (< t)$ are required as the initial value. 
The initial value is not optimized in RC. 
Instead, we employ $f$, which reduces the dependence on the past input through the recursive equation Eq.~(\ref{eq:reservoir_nodes}), and take a sufficiently large $\tau$ to decrease the initial value dependence. 
For example, the hidden node values of the ESN, a typical example of RC systems, are given by~\cite{jaeger2002tutorial} 
\begin{align}
 x_t =& \tanh(v_1^\top x_{t-1} + v_2^\top u_{t-1} + c).
 \label{eq:ESN_nodes}
\end{align}

The cost function for an RC system is usually given by the MSE between the reservoir output sequence $(y_1, \ldots, y_T)$ and the target output sequence $(\hat{y}_1, \ldots, \hat{y}_T)$ for the input sequence $(\ldots, u_1, \ldots, u_T)$, 
\footnote{Although not used in this paper, a regularization term for $w$ may be added to the cost function.}
\begin{align}
 \frac{1}{T} \sum_{t=1}^T \|\hat{y}_t - y_t\|^2 = \frac{1}{T} \sum_{t=1}^T \|\hat{y}_t - w^\top x_t\|^2. 
 \label{eq:RC_cost}
\end{align}
The linear transformation of the readout layer, $w$, is optimized by 
\begin{align}
 w_T = \argmin_w \frac{1}{T} \sum_{t=1}^T \|\hat{y}_t - w^\top x_t\|^2.
 \label{eq:RC_optimize}
\end{align}
In addition to the MSE, a quantity called the IPC, defined below, 
\begin{align}
 C_T = 1 - \frac{\min_w \frac{1}{T} \sum_{t=1}^T \|\hat{y}_t - w^\top x_t\|^2}{\frac{1}{T} \sum_{t=1}^T \|\hat{y}_t\|^2},
 \label{eq:finite_IPC}
\end{align}
is also used as a performance index for RC. 
Due to the normalization by the target output, the IPC ranges from $0$ to $1$, representing the accuracy of the approximation. 
Since the purpose of RC is to learn the relationship between input and output time series from a finite-length input/output set (training data), the actual performance is given by $\lim_{T \to \infty} C_T$. 
Therefore, we should estimate the IPC for infinitely long training data from the IPC for finite-length training data. 
We call $C_0 \equiv \lim_{T \to \infty} C_T$ the true IPC to distinguish it from the training IPC for finite-length data. 
The simplest estimation method is to use sufficiently long training data. 
Furthermore, previous studies have introduced a threshold to Eq.~(\ref{eq:finite_IPC}) to avoid overestimating the IPC, which takes $0$ in infinite time limit~\cite{dambre2012information, kubota2021unifying, vettelschoss2021information, schulte2023refined, takasu2025neuronal}. 
In this paper, we call their IPC the empirical IPC defined by 
\begin{align}
 C_T^{\rm emp} = \theta(C_T - C_{T, {\rm th}}) \, C_T, 
 \label{eq:IPC_emp}
\end{align}
where $\theta$ is the Heaviside step function and $C_{T, {\rm th}}$ is the threshold determined by $\chi^2$ distribution~\cite{dambre2012information} or the random shuffle surrogate~\cite{kubota2021unifying}. 
This method has the problem that we cannot evaluate the deviation from the limit value when the IPC for infinite data length is nonzero.

\section{Evaluation of the IPC by the asymptotic expansion}
\subsection{Asymptotic expansion of the IPC}
In this section, we estimate the IPC for infinitely long data using the asymptotic expansion. 
First, we summarize the RC dynamics by referring to Refs.~\cite{dambre2012information, grigoryeva2018echo, gonon2019reservoir}. 
From Eq.~(\ref{eq:reservoir_nodes}), the values of the hidden nodes of the RC system at time $t$ are determined by the initial value $x_{-\tau}$ and the input sequence $(u_{-\tau}, \ldots, u_t)$. 
As we have explained, the RC system is usually designed to reduce dependence on the initial value and the past input. 
Hence, taking $\tau \to \infty$, we can consider that $x_t$ is determined only by the input sequence $(u_{t - s})_{s=0}^\infty$. 
Due to time-independence of Eq.~(\ref{eq:reservoir_nodes}), if we give the same input sequence at two different times, $(u_{t-s})_{s=0}^\infty = (v_{t'-s})_{s=0}^\infty$, the hidden node values at $t$ with the input sequence $(u_{t-s})_{s=0}^\infty$ and those at $t'$ with $(v_{t'-s})_{s=0}^\infty$ are the same. 
This means that the values of the hidden nodes are determined by a sequence of real numbers, and the hidden nodes of the RC system can be considered as the mapping from a real number sequence (input sequence) to a real vector (hidden node variables). 
To simplify the notation, in the following, the input sequence $(u_{t - s})_{s=0}^\infty$ is represented as $u'_t$, and ${\cal U} \subseteq (\mathbb{R}^{d_0})^{\mathbb{N}}$ denotes the set of input sequences. 
We express the target output function as $\hat{y} \colon {\cal U} \to \mathbb{R}^{d_2}$ and the reservoir function as $x \colon {\cal U} \to \mathbb{R}^{d_1+1}$. 
We assume that the components of $x$, that is, $x_1, \ldots, x_{d_1}, x_{d_1+1}$, are linearly independent. 

Next, we consider the stochasticity of the input sequence following Ref.~\cite{gonon2019reservoir}. 
Let $U = (U_t)_{t \in \mathbb{Z}}$ be a stochastic process. 
We use the notation $U'_t = (U_s)_{s \leq t}$, and in particular $U' = (U_t)_{t \leq 0}$. 
Since $\hat{y}$ and $x$ are measurable functions in general, the composite functions, $\hat{y} \circ U'_t$ and $x \circ U'_t$, are random variables. 
We divide the target output sequence $\{\hat{y} \circ U'_1, ..., \hat{y} \circ U'_T, ..., \hat{y} \circ U'_{T+T'}\}$ into training data $\{ (U'_t, \hat{y} \circ U'_t) \}_{t=1}^T$ and test data $\{ (U'_t, \hat{y} \circ U'_t) \}_{t=T+1}^{T+T'}$. 
We assume $T' = O(T)$ in this paper. 
The readout layer is optimized by the training data, 
\begin{align}
 w_T(u'_T) =& \min_w \frac{1}{T} \sum_{t=1}^T (\hat{y}(u'_t) - w^\top x(u'_t))^2,
 \label{eq:w_MLE}
\end{align}
where we use the notation $(\hat{y}(u'_t))^2 = \hat{y}(u'_t)^\top \hat{y}(u'_t)$. 
The IPCs for the training and test data at $w_T$ are called the training IPC and the test IPC, respectively, which are given by 
\begin{align} 
 C_T(u'_T) =& 1 - \frac{\frac{1}{T} \sum_{t=1}^T (\hat{y}(u'_t) - w_T(u'_T)^\top x(u'_t))^2}{\frac{1}{T} \sum_{t=1}^T (\hat{y}(u'_t))^2} \nonumber \\
  =& 1 - \frac{\bar{l}_T(w_T(u'_T), u'_T)}{\bar{\mu}_T(u'_T)}, 
  \label{eq:IPC_train} \\ 
 C'_{T, T'}(u'_{T+T'}) =& 1 - \frac{\frac{1}{T'} \sum_{t=T+1}^{T+T'} (\hat{y}(u'_t) - w_T(u'_T)^\top x(u'_t))^2}{\frac{1}{T'} \sum_{t=T+1}^{T+T'} (\hat{y}(u'_t))^2} \nonumber \\
 =& 1 - \frac{\bar{l}'_{T, T'}(w_T(u'_T), u'_{T+T'})}{\bar{\mu}_{T'}(u'_{T+T'})}. 
 \label{eq:IPC_test}
\end{align}
For later convenience, we express the numerators and denominators of Eqs.~(\ref{eq:IPC_train}) and (\ref{eq:IPC_test}) as 
\begin{align}
 \bar{l}_T(w, u'_T) =& \frac{1}{T} \sum_{t=1}^T (\hat{y}(u'_t) - w^\top x(u'_t))^2,
 \label{eq:reservoir_norm_train} \\
 \bar{l}'_{T, T'}(w, u'_{T+T'}) =& \frac{1}{T'} \sum_{t=T+1}^{T+T'} (\hat{y}(u'_t) - w^\top x(u'_t))^2,
 \label{eq:reservoir_norm_test} \\
 \bar{\mu}_T(u'_T) =& \frac{1}{T} \sum_{t=1}^T (\hat{y}(u'_t))^2.
 \label{eq:teacher_norm}
\end{align}

Now, we derive the asymptotic forms of the training and the test IPC. 
As defined in Section 2, the IPC for infinitely long training data is called the true IPC $C_0$, and $w$ at that time is called the true parameter $w_0$. 
Since RC and the optimized value $w_T$ can be considered a linear regression model and its optimal parameter, the asymptotic theory of linear regression models in Ref.~\cite{white2014asymptotic} can be applied. 
First, we introduce the notation, 
\begin{align} 
 l(w, u'_t) =& (\hat{y}(u'_t) - w^\top x(u'_t))^2, \\
 \mu(u'_t) =& \hat{y}(u'_t)^2, 
\end{align}
and define the following quantities; 
\begin{align}
 I_\infty(w) 
 =& \frac{1}{4} \sum_{t=0}^\infty {\rm Cov}(\nabla_w l(w, U'_0), \nabla_w l(w, U'_t)^\top) \nonumber \\
 &+ \frac{1}{4}\sum_{t=1}^\infty {\rm Cov}(\nabla_w l(w, U'_0), \nabla_w l(w, U'_{-t})^\top), 
  \label{def:I_inf} \\ 
 J =& \frac{1}{2} \frac{\partial^2 \mathbb{E}[l(w, U')]}{\partial w \partial w}\biggr|_{w = w_0}, \ \ 
 J_{ij, kl} = \mathbb{E}[x_i(U') \, x_k(U') \, \delta_{jl}], 
  \label{def:J} \\
 V_{\mu, \infty} 
  =& \sum_{t=0}^\infty {\rm Cov}(\mu(U'_0), \mu(U'_t)) 
  + \sum_{t=1}^\infty {\rm Cov}(\mu(U'_0), \mu(U'_{-t})), 
  \label{def:V_mu_inf} \\
 C_{l, \mu}(w) 
  =& \sum_{t=0}^\infty {\rm Cov}(l(w, U'_0), \mu(U'_t)) 
  + \sum_{t=1}^\infty {\rm Cov}(l(w, U'_0), \mu(U'_{-t})), 
  \label{def:l_mu_cov} \\ 
 V_{l, \infty}(w) 
  =& \sum_{t=0}^\infty {\rm Cov}(l(w, U'_0), l(w, U'_t))
  + \sum_{t=1}^\infty {\rm Cov}(l(w, U'_0), l(w, U'_{-t})). 
  \label{def:l_auto_cov}
\end{align}
We assume that the matrices above are finite. 
We also assume that the central limit theorem (CLT) holds for $l(w, U'_t)$, $\nabla l(w, U'_t)$, and $\mu(U'_t)$, 
\begin{align}
 \frac{1}{\sqrt{T}} \sum_{t=1}^T (l(w, U'_t) - \mathbb{E}[l(w, U')])
  \xrightarrow{d}& {\cal N}(0, V_{l, \infty}(w)),
  \label{eq:l_t_converge} \\
 \frac{1}{\sqrt{T}} \sum_{t=1}^T (\nabla_w l(w, U'_t) - \nabla_w \mathbb{E}[l(w, U')])
  \xrightarrow{d}& {\cal N}(0, I_\infty(w)),
  \label{eq:nabla_l_t_converge} \\
 \frac{1}{\sqrt{T}} \sum_{t=1}^T (\mu(U'_t) - \mathbb{E}[\mu(U')])
  \xrightarrow{d}& {\cal N}(0, V_{\mu, \infty}), 
  \label{eq:mu_t_converge}
\end{align}
and the law of large numbers (LLN) holds for $x(U'_t) x(U'_t)^\top$, 
\begin{align}
 \frac{1}{T} \sum_{t=1}^T x(U'_t) x(U'_t)^\top \xrightarrow{p}& J. 
\end{align}
Using the notation $l(w) \equiv \mathbb{E}[l(w, U')]$ and $\mu_0 \equiv \mathbb{E}[\mu(U')]$, we find that $C_0$ and $w_0$ can be expressed by 
\begin{align}
 C_0 =& 1 - \frac{\min_w l(w)}{\mu_0} = 1 - \frac{l(w_0)}{\mu_0},
 \label{eq:true_IPC} \\
 w_0 =& \argmin_w l(w).
 \label{eq:true_parameter}
\end{align}
In the following, we express $I_\infty(w_0), C_{l,\mu}(w_0)$, and $V_{l, \infty}(w_0)$ as $I_\infty, C_{l,\mu}$, and $V_{l, \infty}$, respectively. 
From Eq.~(\ref{eq:nabla_l_t_converge}), the following quantities, 
\begin{align}
 \xi_T 
  =& \frac{1}{2} J^{-1/2} \nabla_w \frac{1}{\sqrt{T}} \sum_{t=1}^T (l(w, U'_t) - l(w)) \Bigr|_{w=w_0} \nonumber \\
  =& \frac{1}{2 \sqrt{T}} J^{-1/2} \sum_{t=1}^T \nabla_w l(w_0, U'_t) \, , \\
 \xi'_{T'} 
  =& \frac{1}{2} J^{-1/2} \nabla_w \frac{1}{\sqrt{T'}} \sum_{t=T+1}^{T+T'} (l(w, U'_t) - l(w)) \Bigr|_{w=w_0} \nonumber \\
  =& \frac{1}{2 \sqrt{T'}} J^{-1/2} \sum_{t=T+1}^{T+T'} \nabla_w l(w_0, U'_t) \, ,
\end{align}
converge in distribution,  
\begin{align}
 \xi_T \xrightarrow{d}& {\cal N}(0, J^{-1/2} I_\infty J^{-1/2}),
  \label{eq:xi_converge} \\
 \xi'_{T'} \xrightarrow{d}& {\cal N}(0, J^{-1/2} I_\infty J^{-1/2}). 
  \label{eq:xi'_converge}
\end{align}
Following Ref.~\cite{watanabe2009algebraic}, we expand $\nabla_w \bar{l}_T(w_T(U'_T), U'_T)$ around $w_0$ and obtain 
\begin{align}
 0 =& \nabla_w \bar{l}_T(w_T(U'_T), U'_T) = \nabla_w \bar{l}_T(w_0, U'_T) + \nabla_w^2 \bar{l}_T(U'_T) \, (w_T(U'_T) - w_0), 
 \label{eq:MVT_MLE_params} \\
 \nabla_w \bar{l}_T(w_0, U'_T) 
  =& - \frac{2}{T} \sum_{t=1}^T (\hat{y}(U'_t) - w_0^\top x(U'_t)) \, x(U'_t)^\top, \\
 \nabla_w^2 \bar{l}_T(U'_T) =& \frac{2}{T} \sum_{t=1}^T x(U'_t) \, x(U'_t)^\top. 
\end{align}
Substituting $\nabla_w l_T(w_0, U'_T) = 2 (J / T)^{1/2} \xi_T$ and $\nabla_w^2 \bar{l}_T(U'_T) = 2 J + o_p(1)$ into Eq.~(\ref{eq:MVT_MLE_params}), we find the asymptotic form of $w_T$, 
\begin{align}
 w_T(U'_T) = w_0 - (T J)^{-1/2} \, \xi_T + o_p\left( \frac{1}{\sqrt{T}}\right).
  \label{eq:MLE_params_convergence}
\end{align}
Hence, expanding Eq.~(\ref{eq:reservoir_norm_train}) around $w_T(U'_T)$, we find that 
\begin{align}
 \bar{l}_T(w_0, U'_T)
 =& \bar{l}_T(w_T(U'_T), U'_T) + \nabla_w {\bar{l}_T}(w_T(U'_T), U'_T)^\top (w_0 - w_T) \nonumber \\
 &+ \frac{1}{2} (w_0 - w_T(U'_T))^\top \, \nabla_w^2 \bar{l}_T(U'_T) \, (w_0 - w_T(U'_T)). 
  \label{eq:MVT_train_error}
\end{align}
Substituting $\nabla_w \bar{l}_T(w_T(U'_T), U'_T) = 0$, $\nabla_w^2 \bar{l}_T(U'_T) = 2 J + o_p(1)$ and Eq.~(\ref{eq:MLE_params_convergence}) into Eq.~(\ref{eq:MVT_train_error}), we obtain 
\begin{align}
 \bar{l}_T(w_T(U'_T), U'_T) = \bar{l}_T(w_0, U'_T) - \frac{1}{T} \xi_T^2 + o_p\left(\frac{1}{T}\right). 
 \label{eq:train_error_asymp}
\end{align}
Similarly, expanding Eq.~(\ref{eq:reservoir_norm_test}) around $w_0$, we find that 
\begin{align}
 \bar{l}'_{T, T'}(w_T(U'_T), U'_{T+T'})
 =& \bar{l}'_{T, T'}(w_0, U'_{T+T'}) + \nabla_w \bar{l}'_{T, T'}(w_0, U'_{T+T'})^\top (w_T(U'_T) - w_0) \nonumber \\
 &+ \frac{1}{2} (w_T(U'_T) - w_0)^\top \, \nabla_w^2 \bar{l}'_{T, T'}(U'_{T+T'}) \, (w_T(U'_T) - w_0). 
  \label{eq:MVT_test_error} 
\end{align}
Substituting $\nabla_w \bar{l}'_{T, T'}(w_0, U'_{T+T'}) = 2(J / T')^{1/2} \xi'_{T'}$, 
$\nabla_w^2 \bar{l}'_{T, T'}(U'_{T+T'}) = 2 J + o_p(1)$ and Eq.~(\ref{eq:MLE_params_convergence}) into Eq.~(\ref{eq:MVT_test_error}), we obtain 
\begin{align} 
 \bar{l}'_{T, T'}(w_T(U'_T), U'_{T+T'}) 
  =& \bar{l}'_{T, T'}(w_0, U'_{T+T'}) + \frac{1}{T} (\xi_T)^2 + \frac{2}{\sqrt{T T'}} \, \xi_T^\top \, \xi'_{T'} + o_p\left( \frac{1}{T} \right) \, . 
  \label{eq:test_error_asymp} 
\end{align}
Now, substituting Eqs.~(\ref{eq:train_error_asymp}), (\ref{eq:test_error_asymp}) and (\ref{eq:mu_t_converge}) into Eqs.~(\ref{eq:IPC_train}) and (\ref{eq:IPC_test}), we obtain the asymptotic forms of the training and the test IPC, 
\begin{align} 
 & C_T(U'_T) \nonumber \\
 =& \left(1 - \frac{\bar{l}_T(w_0, U'_T)}{\bar{\mu}_T(U'_T)}\right) + \frac{1}{T} \frac{(\xi_T)^2}{\bar{\mu}_T(U'_T)} + o_p\left(\frac{1}{T}\right)
  \label{eq:asymp_error_train_IPC} \\
 =& 1 - \frac{1}{\mu_0} \left[ \bar{l}_T(w_0, U'_T) - \frac{\bar{l}_T(w_0, U'_T) \zeta_T}{\mu_0} + \frac{\bar{l}_T(w_0, U'_T) \zeta_T^2}{\mu_0^ 2} - \frac{(\xi_T)^2}{T}\right] + o_p\left(\frac{1}{T}\right), 
  \label{eq:asymp_train_IPC} \\ 
 & C'_{T,T'}(U'_{T+T'}) \nonumber \\
 =& \left(1 - \frac{\bar{l}'_{T, T'}(w_0, U'_{T+T'})}{\bar{\mu}_{T'}(U'_{T+T'})} \right) - \frac{1}{T} \left(  \frac{(\xi_T)^2}{\bar{\mu}_{T'}(U_{T+T'})} + \sqrt{\frac{T}{T'}} \frac{2 \xi_T^\top \xi'_{T'}} {\bar{\mu}_{T'}(U'_{T+T'})} \right) + o_p\left(\frac{1}{T}\right) 
 \label{eq:asymp_error_test_IPC} \\
 =& 1 - \frac{1}{\mu_0} \Biggl[ \bar{l}'_{T, T'}(w_0, U'_{T+T'}) - \frac{\bar{l}'_{T, T'}(w_0, U'_{T+T'}) \zeta_{T'}}{\mu_0} + \frac{\bar{l}'_{T'}(w_0, U'_{T+T'}) \zeta_{T'}^2}{\mu_0^2} \nonumber \\
 &+ \frac{(\xi_T)^2}{T} + \frac{2 \xi_T^\top \xi'_{T'}} {\sqrt{T T'}} \Biggr] + o_p\left(\frac{1}{T}\right), 
 \label{eq:asymp_test_IPC} \\
 \zeta_T \equiv& \bar{\mu}_T(U'_T) - \mu_0, \ \  \zeta_{T'} \equiv \bar{\mu}_{T'}(U'_{T+T'}) - \mu_0.
\end{align}
Note that we have assumed $T' = O(T)$. 
The first and the rest terms of Eq.~(\ref{eq:asymp_error_train_IPC}) represent the model assumption bias and the finite-size effect, respectively. 
If the RC system can reproduce the target function $\hat{y}$, the first term converges to $1$ in $T \to \infty$. 
The first and the rest terms of Eq.~(\ref{eq:asymp_error_test_IPC}) can similarly be interpreted. 

Taking expectations for the training and the test data in Eqs.~(\ref{eq:asymp_train_IPC}), the mean and the variance of the training IPC are given by 
\begin{align} 
 \mathbb{E}[C_T(U'_T)] 
  =& 1 - \frac{l(w_0)}{\mu_0} + \frac{1}{T} \left[ \frac{C_{l, \mu}}{\mu_0^2} - \frac{l(w_0) \, V_{\mu, \infty}}{\mu_0^3} + \frac{{\rm Tr}(I_\infty J^{-1})}{\mu_0} \right] + o\left( \frac{1}{T} \right) \, , 
  \label{eq:asymp_train_IPC_mean} \\ 
 \mathbb{V}[C_T(U'_T)] 
  =& \frac{1}{T} \left[ \frac{V_{l, \infty}}{\mu_0^2} + \frac{l(w_0)^2 \, V_{\mu, \infty}}{\mu_0^4} - \frac{2}{\mu_0^3} \, l(w_0) \, C_{l, \mu} \right] + o\left( \frac{1}{T} \right) \, .
 \label{eq:asymp_train_IPC_var} 
\end{align}
Similarly, from Eq.~(\ref{eq:asymp_test_IPC}), the mean and variance of the test IPC are given by
\begin{align} 
 \mathbb{E}[C'_{T, T'}(U'_{T+T'})] 
  =& 1 - \frac{l(w_0)}{\mu_0} + \frac{1}{T} \left[ \frac{T}{T'} \, \frac{C_{l, \mu}} {\mu_0^2} - \frac{T}{T'} \, \frac{l(w_0) \, V_{\mu, \infty}}{\mu_0^3} - \frac{{\rm Tr}(I_\infty J^{-1})}{\mu_0} \right] + o\left( \frac{1}{T} \right) \, , 
  \label{eq:asymp_test_IPC_mean} \\ 
 \mathbb{V}[C'_{T, T'}(U'_{T+T'})] 
  =& \frac{1}{T} \left[ \frac{T}{T'} \, \frac{V_{l, \infty}}{\mu_0^2} + \frac{T}{T'} \, \frac{l(w_0)^2 \, V_{\mu, \infty}}{\mu_0^4} - \frac{T}{T'} \, \frac{2}{\mu_0^3} \, l(w_0) \, C_{l, \mu} \right] + o\left( \frac{1}{T} \right) \, .
  \label{eq:asymp_test_IPC_var}
\end{align}
From Eqs.~(\ref{eq:asymp_train_IPC_mean}) and (\ref{eq:asymp_test_IPC_mean}), we find that both means approach the true IPC, $C_0 = 1 - l(w_0) / \mu_0$, in $O(1/T)$. 
To estimate the true IPC, we first approximate the expectation values, $\mathbb{E}[C_T(U'_T)]$ and $\mathbb{E}[C'_{T, T'}(U'_{T+T'})]$, by the sample mean at each $T$. 
Then we ignore the term $o(1/T)$ in Eqs.~(\ref{eq:asymp_train_IPC_mean}) and (\ref{eq:asymp_test_IPC_mean}) and use the weighted least squares method to estimate the true IPC (see Appendix). 

Note that fitting the variances by Eqs.~(\ref{eq:asymp_train_IPC_var}) and (\ref{eq:asymp_test_IPC_var}) fails when the true IPC is $0$. 
In this case, the true parameter is $w_0 = 0$, and we find $l(w_0, U') = (\hat{y}(U'))^2$. 
Thus, we obtain 
\begin{align}
 & {\rm Cov}(l(w_0, U'_0), l(w_0, U'_t)) = {\rm Cov}(l(w_0, U'_0), (\hat{y}(U'_t))^2) \nonumber \\
 =& {\rm Cov}(l(w_0, U'_0), \mu(U'_t)) = {\rm Cov}(\mu(U'_0) , \mu(U'_t)). 
\end{align}
That is, $V_{l, \infty} = V_{\mu, \infty} = C_{l, \mu}$. 
Since the coefficients of the $1/T$ terms in Eqs.~(\ref {eq:asymp_train_IPC_var}) and (\ref{eq:asymp_test_IPC_var}) vanish, fitting the coefficients of the $1/T$ terms fails. 
Although we do not show whether the converse holds, we use this property to
determine whether the true IPC is $0$. 
If the estimated true IPC is close to $0$ and the power of the variance is not $1/T$, we propose that the true IPC is taken to be $0$. 
In Ref.~\cite{dambre2012information}, by assuming that $x(U_t)$ and $\hat{y}(U_t)$ are independent at each time, it is derived that when the true IPC is $0$, the variance of the training IPC at $T$ is $O(1/T^2)$, which is consistent with our result. 
Since discrimination by the power of the variance is less affected by fluctuations in finite-length data than discrimination by numerical value, our method is expected to be more robust. 
 

\subsection{Remarks on the proposed method}
We make three comments. First, when we evaluate the training and the test IPC, we need not store $(x(u'_t))_{t=1}^{T+T'}$ and $(\hat{y}(u_t))_{t=1}^{T+T'}$. 
Since Eqs.~(\ref{eq:w_MLE}), (\ref{eq:IPC_train}), and (\ref{eq:IPC_test}) can be rewritten as 
\begin{align}
 & w_T(u'_T) 
  = \left< x x^\top \right>_T^{-1} \left< x \hat{y}^\top \right>_T , \\
 & C_T(u'_T) = \frac{1}{\left< \hat{y}^2\right>_T} {\rm Tr}\left[ \left< \hat{y} x^\top\right>_T \left< x x^\top\right>_T^{-1} \left< x \hat{y}^\top\right>_T \right], 
  \label{eq:IPC_train_inverse} \\
 & C'_{T, T'}(u'_{T+T'}) = \frac{1}{\left< \hat{y}^2\right>_{T'}} {\rm Tr}\biggl[ \left< \hat{y} x^\top\right>_T \left< x x^\top\right>_T^{-1} 
  \left( 2 \left< x \hat{y}^\top \right>_{T'} - \left< x x^\top\right>_{T'} \left< x x^\top\right>_T^{-1} \left< x \hat{y}^\top\right>_T \right) \biggr], 
  \label{eq:IPC_test_inverse} \\
 & \left< \hat{y}^2 \right>_T \equiv \frac{1}{T} \sum_{t=1}^T \hat{y}(u'_t)^2, \ 
 \left< \hat{y} x^\top \right>_T \equiv \frac{1}{T} \sum_{t=1}^{T} \hat{y}(u'_t) x(u'_t)^\top, \ 
 \left< x x^\top \right>_T \equiv \frac{1}{T} \sum_{t=1}^{T} x(u'_t) x(u'_t)^\top, 
 \label{eq:summand_1} \\
 & \left< \hat{y}^2 \right>_{T'} \equiv \frac{1}{T'} \sum_{t=T+1}^{T+T'} \hat{y}(u'_t)^2, \ 
 \left< \hat{y} x^\top \right>_{T'} \equiv \frac{1}{T'} \sum_{t=T+1}^{T+T'} \hat{y}(u'_t) x(u'_t)^\top, \
 \left< x x^\top \right>_{T'} \equiv \frac{1}{T'} \sum_{t=T+1}^{T+T'} x(u'_t) x(u'_t)^\top,
 \label{eq:summand_2}
\end{align}
it is only necessary to sequentially store the summand of Eqs.~(\ref{eq:summand_1}) and (\ref{eq:summand_2}), e.g. $\sum_{t=1}^s \hat{y}(u'_t)^2$ at time $s$, at each time up to $T$ or $T + T'$. 
The number of components of the sequences, $(x(u'_t))_{t=1}^{T+T'}$ and $(\hat{y}(u_t))_{t=1}^{T+T'}$, is $(T + T') (d_1 + d_2)$, 
and the number of components of the summand is $2 (d_1^2 + d_1 d_2 + 1)$. 
To estimate the IPC for infinitely long data, $T + T' > 2 d_1$ is satisfied in most cases, and therefore storing the summand instead of the sequences reduces memory consumption. 

Secondly, we estimate the computational complexity of our method. 
The length of data should be longer than the number of parameters, $T > d_1$, and the number of hidden nodes is set to be larger than the input and output dimensions in general, $d_1 \geq d_0, d_2$. 
We assume that the training and test data are of comparable length. 
First, we estimate the computational complexity of the training and the test IPC for each $T$ and $T'$. 
The computational complexity of $x_t$ in Eq.~(\ref{eq:reservoir_nodes}) and summand in Eqs.~(\ref{eq:summand_1}) and (\ref{eq:summand_2}) is $O(d_1^2)$. 
Since we use these equations from $t = 1$ to $t = T + T'$, the computational complexity of the RC dynamics is $O(d_1^2 (T + T')) = O(d_1^2 T)$. 
From Eqs.~(\ref{eq:IPC_train_inverse}) and (\ref{eq:IPC_test_inverse}), the computational complexity of the numerator of the IPC is $O(d_1^3)$. 
Therefore, the computational complexity of a single IPC calculation is $O(d_1^2 T)$. 
If we evaluate the IPC $N$ times to obtain the sample mean, the computational complexity is $O(N d_1^2 T)$. 
As we will empirically see in the next section, $T$ and $T'$ should be more than $10d_1$. 
If the data length is taken from $10 d_1$ to $100 d_1$ for every $10d_1$, the total computational complexity is approximately $10^3$ times $O(N d_1^3)$. 
Therefore, if $d_1$ is increased, it will be difficult to calculate as is. 
Since the amount of calculation is a trade-off with the estimation accuracy, we first obtain the asymptotic expansion of the variance of the IPC using short data, then determine $N$ and $T$ required for the desired accuracy to prevent excessive calculations. 
Furthermore, the computation time can be reduced using the bootstrap or jackknife method to decrease the number of samples calculated or by performing parallel computation. 
However, the computation time cannot be drastically reduced, so $d_1 \sim 10^3$ is expected to be the limit. 


Thirdly, our method can be applied to RC systems that satisfy the assumption that the CLT holds for $l(w, U'_t)$, $\nabla l(w, U'_t)$, and $\mu(U'_t)$, the LLN holds for $x(U'_t) x(U'_t)^\top$, and $C_{l,\mu}(w)$ is a finite matrix. 
Thus, for example, if $l(w, U'_t)$, $\nabla_w l(w, U'_t)$, or $\mu(U'_t)$ are not ergodic processes, or the mean or covariance matrices are infinite or time-dependent, then Eqs.~(\ref{eq:asymp_train_IPC}) and (\ref{eq:asymp_test_IPC}) cannot be used.

\section{Numerical simulation}
First, we show the effectiveness of our method in a system where each term of the asymptotic forms except for $o(1/T)$ can be precisely calculated. 
Next, we estimate the true IPCs in systems where the target outputs are given by Legendre polynomials. 
Finally, we apply our method to the estimation of the true IPC for the NARMA10 task.
In these simulations, we compared the proposed method with the method in Ref.~\cite{dambre2012information}. 
The threshold in Eq.~(\ref{eq:IPC_emp}) is given by $\chi^2$ distribution as 
\begin{align}
 C_{T, {\rm th}} = \frac{2 \alpha}{T}, \ \ {\rm Prob}(\chi^2(N) \geq \alpha) = p,
\end{align}
where $p$ is a small positive real number, and we set $p=10^{-4}$. 

 \subsection{A simple model}
In this model, $U_t$ is a random variable from ${\rm Uniform}(-1, 1)$. 
The hidden node of the RC system and the target output are given by 
\begin{align}
 x(u'_t) = \sum_{s=0}^\infty 2^{-s} \, u_{t-s}, \ \
 \hat{y}(u'_t) = 1 + x(u'_t).
\end{align}
The MSE is 
\begin{align}
 l(w) = \mathbb{E}[( \hat{y}(u') - w \, x(u'))^2] = \frac{4}{9} \, (w-1)^2 + 1 . 
\end{align}
Thus, the true parameter is $w_0 = 1$, and the minimum value of the MSE is $l(w_0) = 1$. 
The true IPC is 
\begin{align}
 C_0 = 1 - \frac{l(w_0)}{\mu_0} = \frac{4}{13}. 
\end{align}
The other terms in the asymptotic expansions are 
\begin{align}
C_{l, \mu} = 0, \ \ V_{l, \infty} = 0, \ \ V_{\mu, \infty} = \frac{6992}{1215}, \ \
I_\infty = \frac{4}{3}, \ \ J = \frac{4}{9}. 
\end{align}

\begin{figure}[htbp]
  \begin{tabular}{cc}
    \begin{minipage}[t]{0.4\hsize}
      \centering
      \includegraphics[keepaspectratio, scale=0.4]{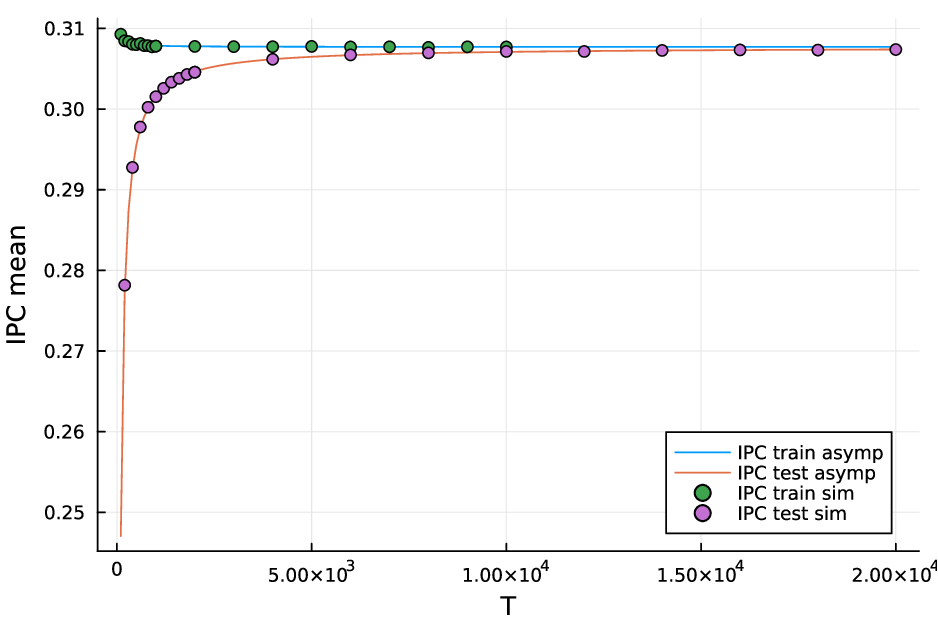}
      \subcaption{Mean of IPCs in the simple model}
      \label{fig:small_sys_IPC_mean}
    \end{minipage} & 
    \begin{minipage}[t]{0.4\hsize}
      \centering
      \includegraphics[keepaspectratio, scale=0.4]{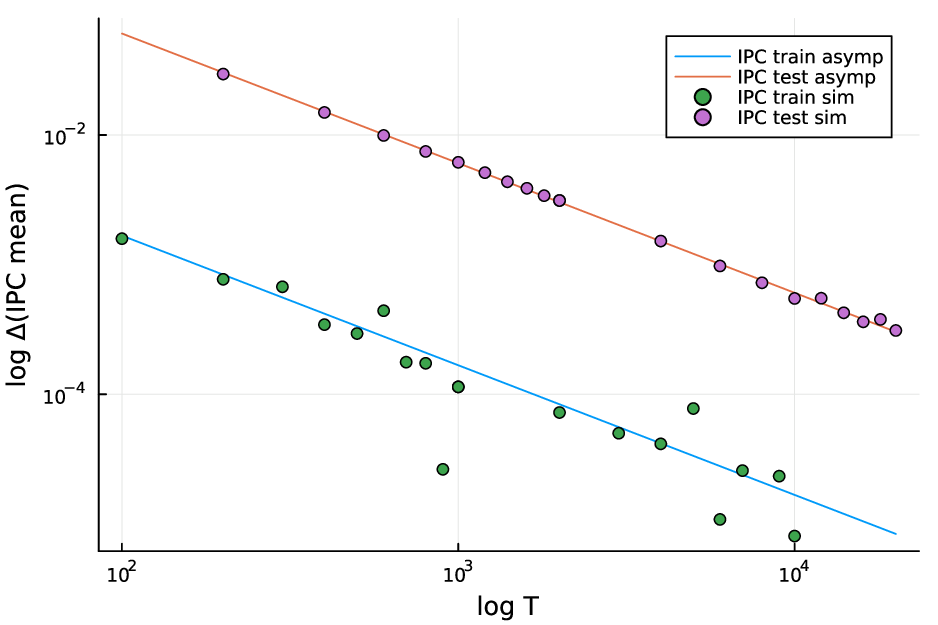}
      \subcaption{Mean of IPCs after removal of the constant term on a log-log scale in the simple model}
      \label{fig:small_sys_IPC_mean_log}
    \end{minipage} \\
    \begin{minipage}[t]{0.4\hsize}
      \centering
      \includegraphics[keepaspectratio, scale=0.4]{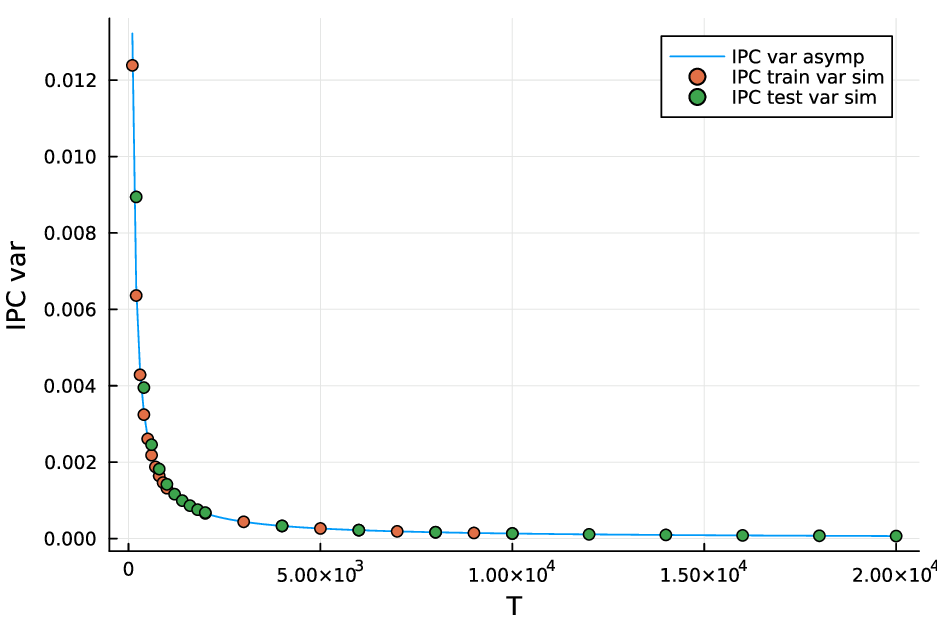}
      \subcaption{Variance of IPCs in the simple model}
      \label{fig:small_sys_IPC_var}
    \end{minipage} &
    \begin{minipage}[t]{0.4\hsize}
      \centering
      \includegraphics[keepaspectratio, scale=0.4]{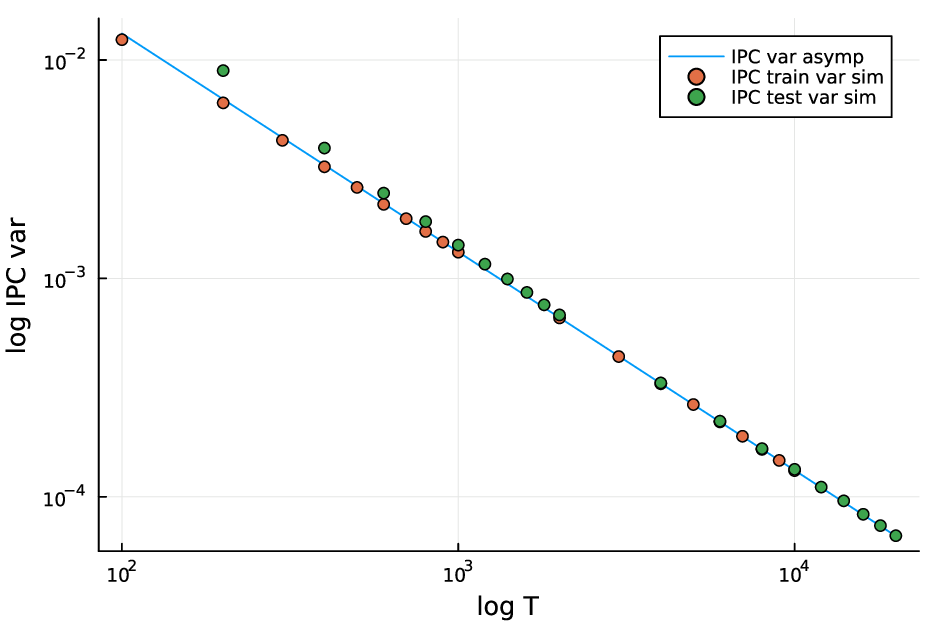}
      \subcaption{Variance of IPCs on a log-log scale in the simple model}
      \label{fig:small_sys_IPC_var_log}
    \end{minipage} \\
  \end{tabular}
  \caption{The means and the variances of the training and the test IPC obtained by simulation in the simple system, and the theoretical lines obtained by the asymptotic expansions are plotted against $T$. Although the means of the training IPC fluctuate in the log-log plot, most samples are roughly on the theoretical lines, which indicates the effectiveness of the estimation.}
\label{fig:small_sys_IPC}
\end{figure}

We performed a numerical simulation. 
Training and test data were generated $10^5$ times for each data length $T$, and the training and the test IPC were calculated. 
Their means and variances were modeled as 
\begin{align}
 C(T) = a + \frac{b_1}{T} , \ \ C'(T, T') = a - \frac{b_2}{T'} , \ \
 V(T) = \frac{d}{T} , \ \ V'(T, T') = \frac{d}{T'},
\end{align}
and $a, b_1, b_2, d$ were estimated using the weighted least squares method. 
To confirm the effectiveness of the asymptotic expansion, we used the first $5$ training and test data, that is, $T = 200, \ldots, 600$ and $T' = 400, \ldots, 1200$. 
The exact and estimated values of the asymptotic expansion are shown in Table~\ref{table:result_simple_model}.
\begin{table}[h]
 \centering
 \caption{The ground truth and the estimated values of the coefficients of the asymptotic forms are shown. We can find that the estimated values matched well to the ground truth.}
 \label{table:result_simple_model}
 \begin{tabular}{|c|c|c|c|c|}
  \hline
    & $a$ & $b_1$ & $b_2$ & $d$ \\ \hline
   ground truth & 4/13 & 1839/10985 & 66606/10985 & 188784/142805 \\ \hline
   estimation & 0.3075 & 0.1933 & 5.8906 & 1.412 \\ \hline
 \end{tabular}
\end{table}
We find that the weighted least-squares estimation is successful. 
We plot the means and variances of the IPC in Fig.~\ref{fig:small_sys_IPC} and find that they are almost on the asymptote up to the term $1/T$. 
Next, we plotted on the log-log scale to verify the dominant $T$-dependence. 
To remove the constant term, we subtracted $C_0$ from the means of the IPCs. 
(After subtracting $C_0$, the 17th item in the mean of the training IPC was negative, and we removed it from the log-log scale.) 
Most of the samples are on the asymptote up to the term $1/T$, which shows the validity of the asymptotic expansion. 
Although the means of the training IPC fluctuate more, this is likely due to a lack of samples. 

Finally, we compare the proposed method with the method in Ref.~\cite{dambre2012information}. 
We used the sample mean of the training IPC at $T = 600$. 
Since the threshold is $C_{T, {\rm th}} = 0.0505$, the empirical IPC is $0.3080$. 
Our result is slightly closer to the ground truth, showing that the proposed method is effective (the differences from the ground truth are $2 \times 10^{-4}$ and $-3 \times 10^{-4}$, respectively).

 \subsection{Legendre polynomials}
An RC system approximates a mapping from input to output time series by the linear combination of nonlinear dynamics. 
To capture the set of functions that RC can produce, it is useful to evaluate the IPCs of the orthogonal polynomials~\cite{dambre2012information, kubota2021unifying}. 
In this subsection, the input sequence $U'$ is sampled independently from ${\rm Uniform}(-1, 1)$, and the target output is the product of Legendre polynomials, 
\begin{align}
\hat{y}(u'_t) = \prod_{i=1}^\infty L_{s_i}(u_{t-i}), 
\end{align}
where, $L_{s_i}$ is the $s_i$-th Legendre polynomial, and the number of nonzero elements in $(s_i)_{i=1}^\infty$ is finite. 

\begin{figure}[htbp]
  \begin{tabular}{cc}
    \begin{minipage}[t]{0.4\hsize}
      \centering
      \includegraphics[keepaspectratio, scale=0.4]{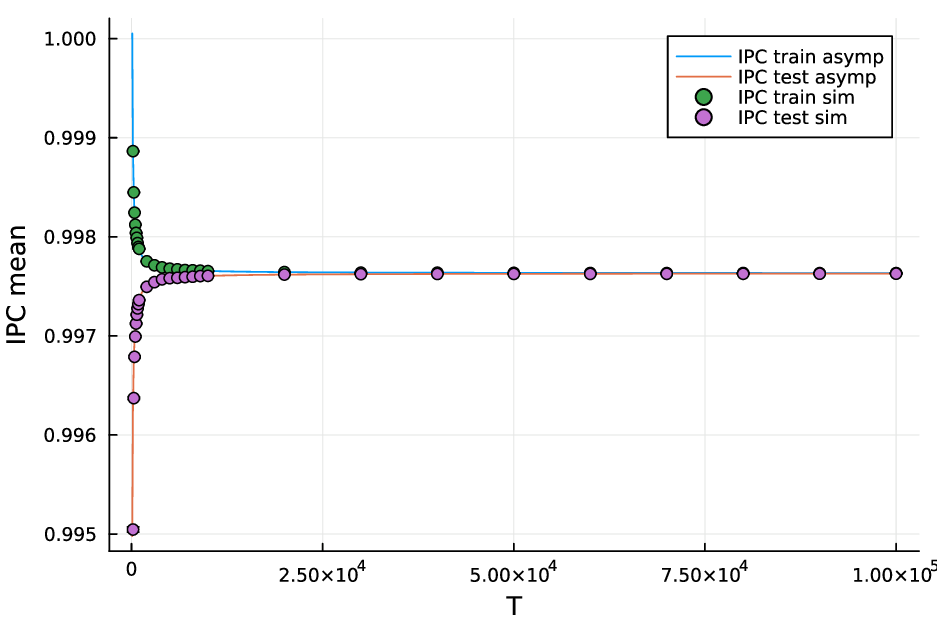}
      \subcaption{Mean of IPCs for the 1st order polynomial task}
      \label{fig:legendre1_IPC_mean}
    \end{minipage} & 
    \begin{minipage}[t]{0.4\hsize}
      \centering
      \includegraphics[keepaspectratio, scale=0.4]{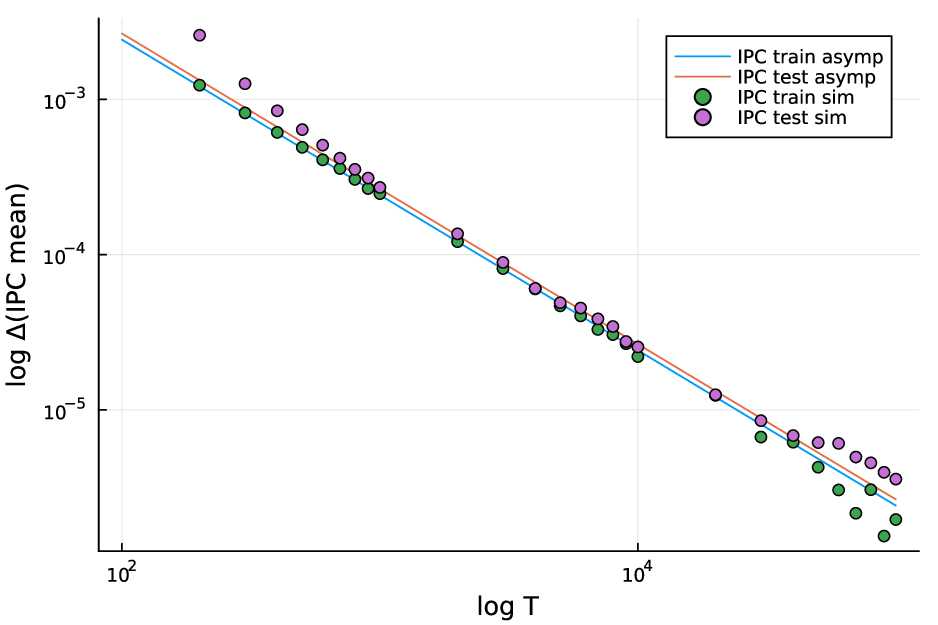}
      \subcaption{Mean of IPCs after removal of the constant term on a log-log scale for the 1st order polynomial task}
      \label{fig:legendre1_IPC_mean_log}
    \end{minipage} \\
    \begin{minipage}[t]{0.4\hsize}
      \centering
      \includegraphics[keepaspectratio, scale=0.4]{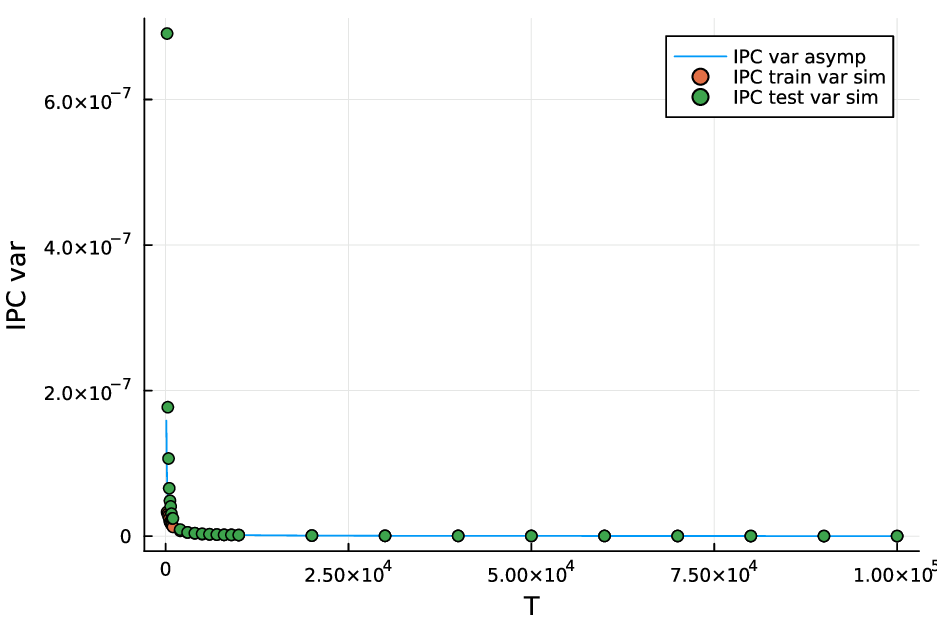}
      \subcaption{Variance of IPCs for the 1st order polynomial task}
      \label{fig:legendre1_IPC_var}
    \end{minipage} &
    \begin{minipage}[t]{0.4\hsize}
      \centering
      \includegraphics[keepaspectratio, scale=0.4]{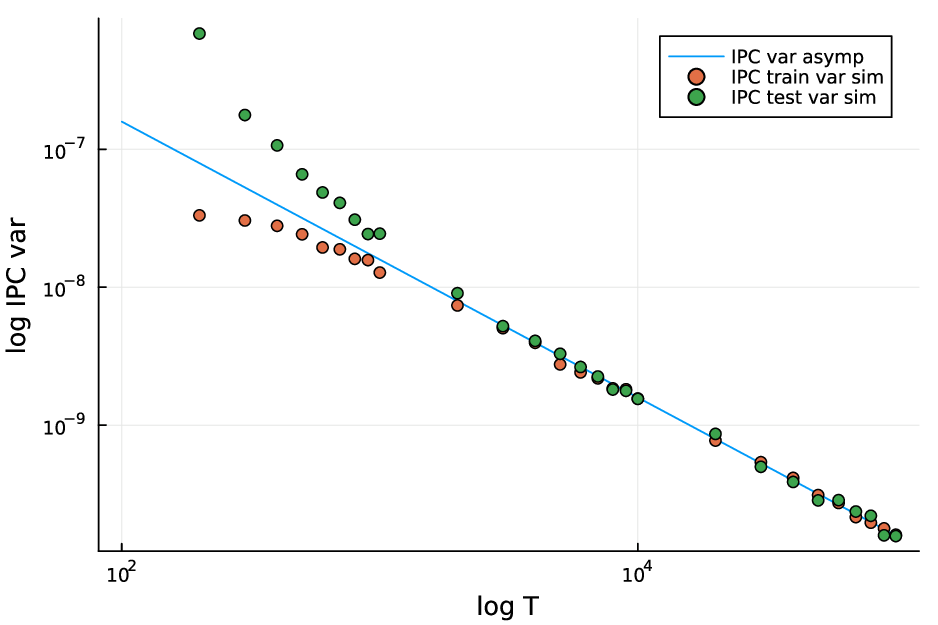}
      \subcaption{Variance of IPCs on a log-log scale for the 1st order polynomial task}
      \label{fig:legendre1_IPC_var_log}
    \end{minipage} \\
  \end{tabular}
  \caption{The means and the variances of the IPC using uniformly distributed input and Legendre first-order output were plotted, along with the asymptote estimated from them. The simulation results were mostly on the theoretical line, indicating the effectiveness of the estimation.}
\label{fig:legendre1_IPC}
\end{figure}

\begin{figure}[htbp]
  \begin{tabular}{cc}
    \begin{minipage}[t]{0.4\hsize}
      \centering
      \includegraphics[keepaspectratio, scale=0.4]{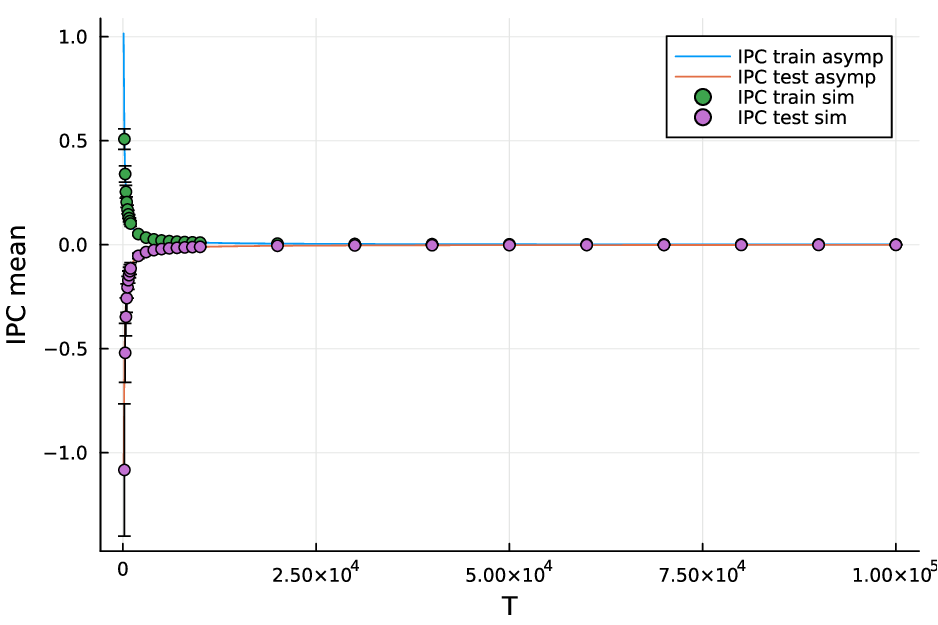}
      \subcaption{Mean of IPCs for the 15th order polynomial task}
      \label{fig:legendre15_IPC_mean}
    \end{minipage} & 
    \begin{minipage}[t]{0.4\hsize}
      \centering
      \includegraphics[keepaspectratio, scale=0.4]{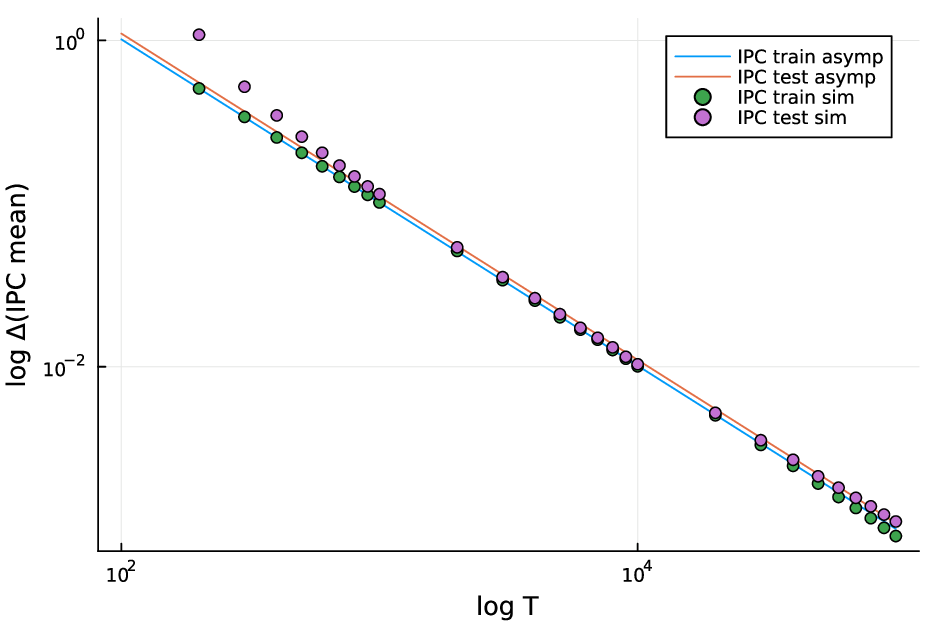}
      \subcaption{Mean of IPCs after removal of the constant term on a log-log scale for the 15th order polynomial task}
      \label{fig:legendre15_IPC_mean_log}
    \end{minipage} \\
    \begin{minipage}[t]{0.4\hsize}
      \centering
      \includegraphics[keepaspectratio, scale=0.4]{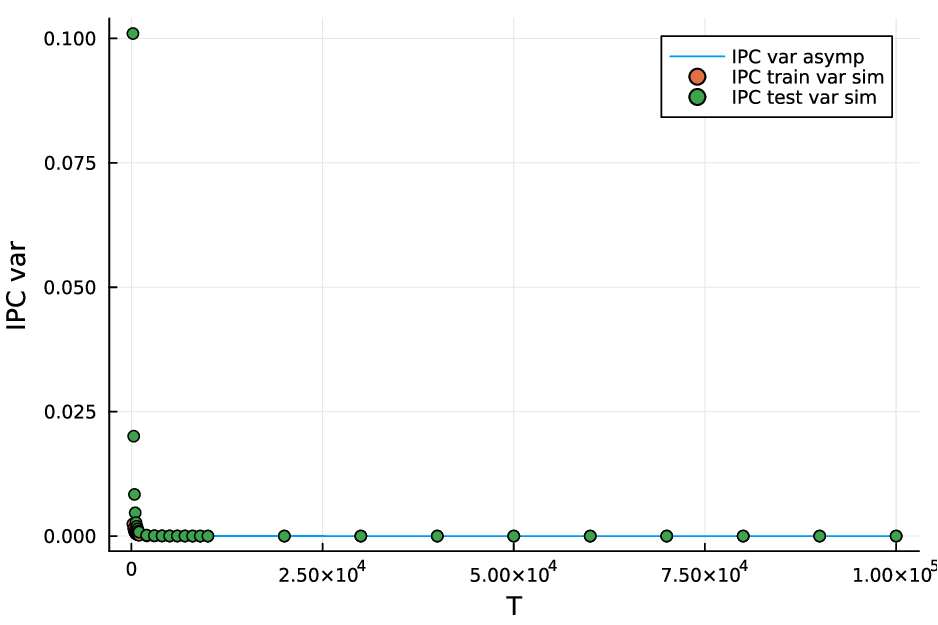}
      \subcaption{Variance of IPCs for the 15th order polynomial task}
      \label{fig:legendre15_IPC_var}
    \end{minipage} &
    \begin{minipage}[t]{0.4\hsize}
      \centering
      \includegraphics[keepaspectratio, scale=0.4]{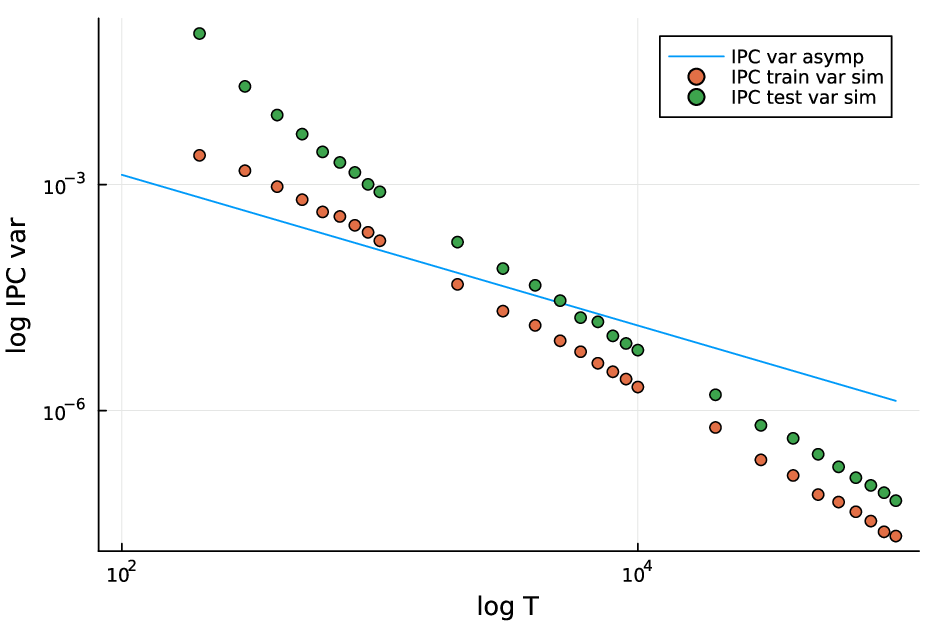}
      \subcaption{Variance of IPCs on a log-log scale for the 15th order polynomial task}
      \label{fig:legendre15_IPC_var_log}
    \end{minipage} \\
  \end{tabular}
  \caption{The means and the variances of the IPC using uniformly distributed input and 15th-order Legendre output are plotted, along with the asymptote estimated from them. Although the means are roughly on the theoretical line, the variances are significantly off. Since the true IPC is nearly $0$, we can find that the variance decays faster than $1/T$. (In this graph, it decays at approximately $1/T^2$.)}
\label{fig:legendre15_IPC}
\end{figure}

We employed two target outputs. 
One was a short-term linear task $\hat{y}_t = L_1(u_{t-1})$, which is easy to approximate with the ESN, and the other was a long-term nonlinear task $\hat{y}_t = L_{15}(u_{t-5})$, which is difficult to approximate. 
The number of hidden nodes was $d_1 = 100$, the ESN parameters, $v_1, v_2$, were randomly chosen from the normal distribution, and $c = 0$. 
The spectral radius of $v_1$ was set to $0.9$, and the proportion of nonzero elements was set to $0.7$. 
For each $T$, we generated data $10^3$ times and calculated the training and the test IPC. 
To confirm the usefulness of the asymptotic expansion, we used the training and test data from $T = 10^3$ to $T = 10^4$ and estimated the asymptotic parameters. 
We obtained $a = 0.997632, b_1 = 0.242030, b_2 = 0.265197, d = 1.6 \times 10^{-5}$ for the first-order task and $a = 0.00070, b_1 = 99.63334, b_2 = 110.70010, d = 0.13467$ for the 15th-order task. 
Figs.~\ref{fig:legendre1_IPC} and \ref{fig:legendre15_IPC} show the results when the first- and 15th-order Legendre polynomials were used for the target output, respectively. 
The means and the variances of the true IPC for the first-order polynomial task are roughly on the asymptotic curve, indicating that the estimation was successful. 
In the 15th-order polynomial task, the asymptote fitted the mean IPC samples well, and the estimated true IPC was almost $0$. 
The variances were significantly off the asymptote in the log-log graph, which reflects our mention in the last paragraph of subsection 3.1 that when the true IPC is $0$, the variance approaches $0$ faster than $1/T$. 

Finally, we compare our method with the method in Ref.~\cite{dambre2012information}. 
We used the sample mean of the training IPC at $T = 10^4$. 
Since the threshold is $C_{T, {\rm th}} = 0.032264$, the empirical IPC for the target output $\hat{y}_t = L_1(u_{t-1})$ is $0.997654$. 
For the target output $\hat{y}_t = L_{15}(u_{t-1})$, the sample mean of the training IPC is $0.01017$, which is below the threshold, and therefore the empirical IPC is $0$. 
Since the ground truth is unknown, it is difficult to compare the results. 
In the first-order task, the proposed method is closer to the IPC $0.997631$ estimated by longer data, $T = 10^5$, which shows the usefulness of the asymptotic expansion. 
In the 15th-order task, both methods produced the same IPC. 
When determining whether the IPC is $0$, it is considered more robust to use the power of the variance, which is less susceptible to fluctuations caused by finite-length data than the numerical value of the IPC.

 \subsection{NARMA10}
\begin{figure}[htbp]
  \begin{tabular}{cc}
    \begin{minipage}[t]{0.4\hsize}
      \centering
      \includegraphics[keepaspectratio, scale=0.4]{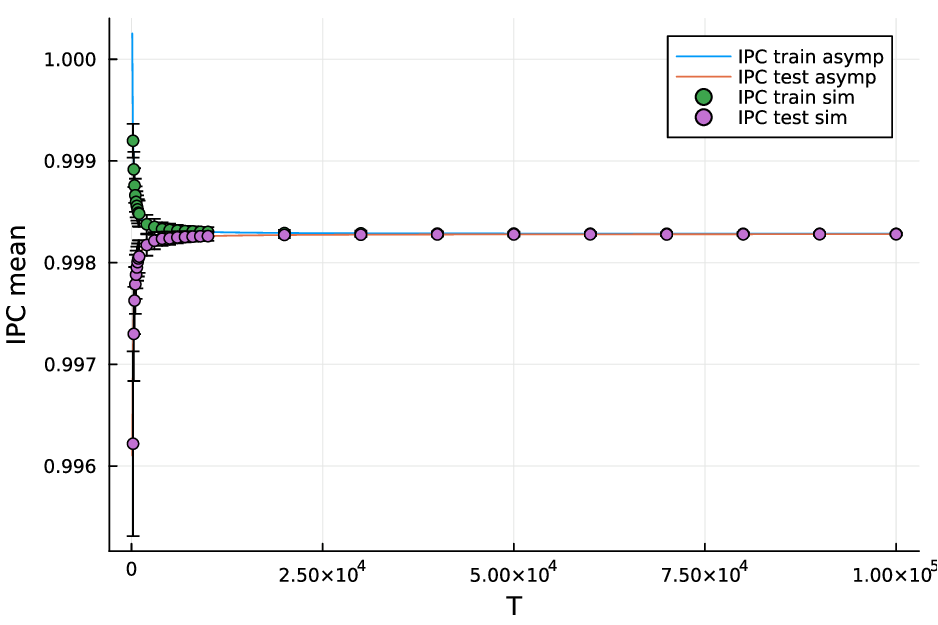}
      \subcaption{Mean of IPCs for the NARMA10 task}
      \label{fig:narma10_IPC_mean}
    \end{minipage} & 
    \begin{minipage}[t]{0.4\hsize}
      \centering
      \includegraphics[keepaspectratio, scale=0.4]{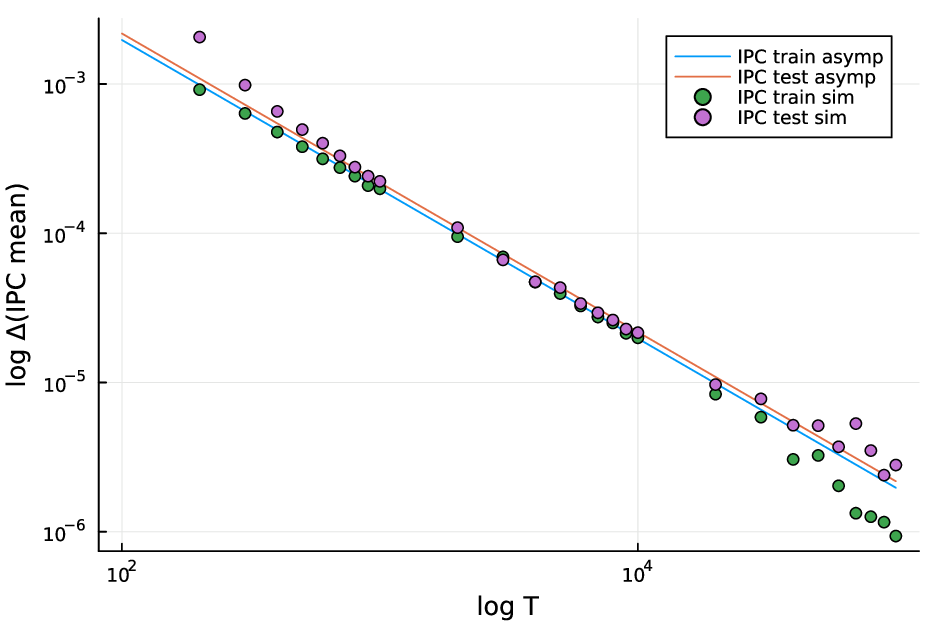}
      \subcaption{Mean of IPCs after removal of the constant term on a log-log scale for the NARMA10 task}
      \label{fig:narma10_IPC_mean_log}
    \end{minipage} \\
    \begin{minipage}[t]{0.4\hsize}
      \centering
      \includegraphics[keepaspectratio, scale=0.4]{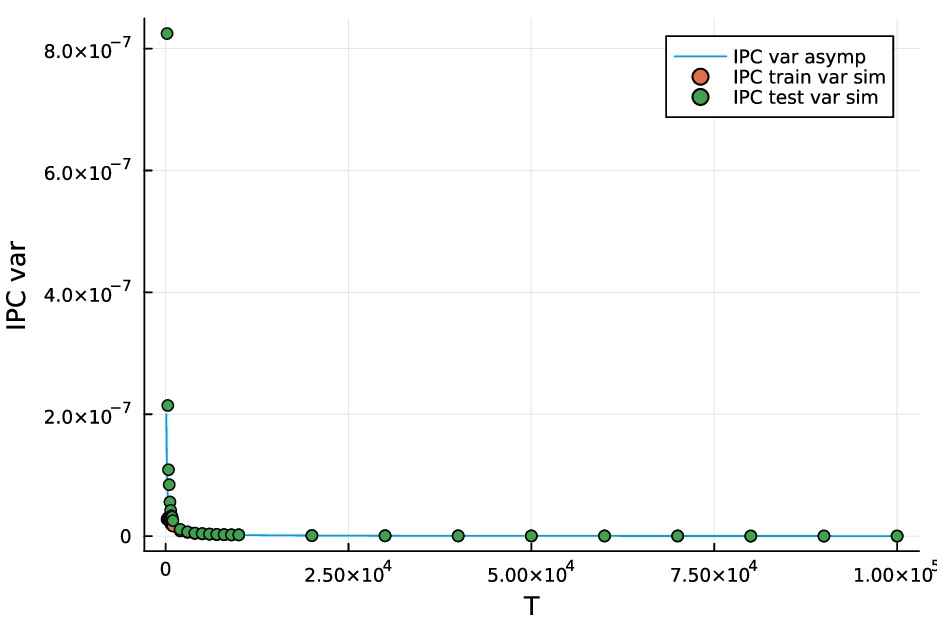}
      \subcaption{Variance of IPCs for the NARMA10 task}
      \label{fig:narma10_IPC_var}
    \end{minipage} &
    \begin{minipage}[t]{0.4\hsize}
      \centering
      \includegraphics[keepaspectratio, scale=0.4]{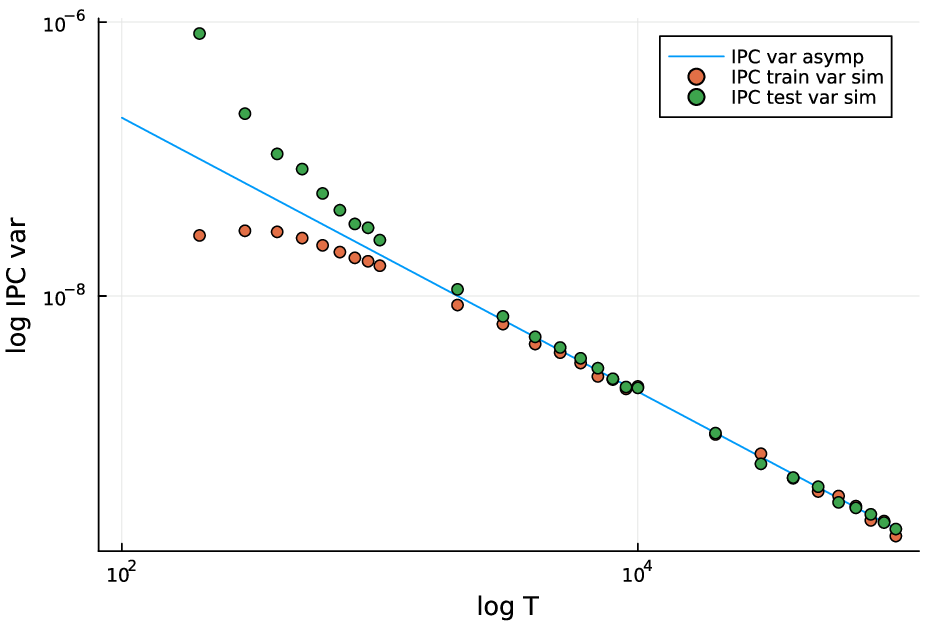}
      \subcaption{Variance of IPCs on a log-log scale for the NARMA10 task}
      \label{fig:narma10_IPC_var_log}
    \end{minipage} \\
  \end{tabular}
  \caption{The means and the variances of the IPC obtained from the NARMA10 task and the asymptote estimated from them are plotted. The simulation results are almost on the theoretical lines.}
\label{fig:narma10_IPC}
\end{figure}

The NARMA10 task is widely used as a benchmark for RC~\cite{atiya2000new}. 
The input sequence is sampled independently from the uniform distribution. 
The output sequence of the NARMA10 is given by 
\begin{align}
 \hat{y}_t 
  = \alpha y_{t-1} + \beta y_{t-1} \sum_{i=1}^{10} y_{t-i} + \gamma u_t u_{t-9} + \delta.
\end{align}
We used ${\rm Uniform}(0, 0.2)$ as the input distribution and $(\alpha, \beta, \gamma, \delta) = (0.3, 0.05, 1.5, 0.1)$ as NARMA10 parameters. 
The number of hidden nodes was $d_1 = 100$, the ESN parameters $v_1, v_2$, were randomly chosen from the normal distribution, and $c = 0$. 
The spectral radius of $v_1$ was set to $0.9$, and the proportion of nonzero elements was set to $0.7$. 
For each $T$, we generated the data $10^3$ times and calculated the training and the test IPC. 
To confirm the usefulness of the asymptotic expansion, we used the training and test data from $T = 10^3$ to $T = 10^4$ and estimated the asymptotic parameters. 
We obtained $a = 0.998283, b_1 = 0.193536, b_2 = 0.215439, d = 2.0 \times 10^{-5}$. 
The results are shown in Fig.~\ref{fig:narma10_IPC}. 
We can find that both the means and variances are almost on the asymptote. 

Finally, we compare our method with the method in Ref.~\cite{dambre2012information}. 
We used the sample mean of the training IPC at $T = 10^4$. 
Since the threshold is $C_{T, {\rm th}} = 0.032264$, the empirical IPC is $0.9983023$. 
Due to the unknown ground truth, it is difficult to compare the results. 
However, the proposed method is closer to the IPC $0.998281$ estimated by longer data, $T = 10^5$, which shows the usefulness of the asymptotic expansion.

\section{Summary}
RNNs and their variants such as LSTM~\cite{hochreiter1997long} and GRU~\cite{cho2014learning} can approximate relationship between input and output time series. 
However, gradient learning takes a long time as the data length is longer. 
RC is a system that approximates the relationship between input and output time series by the linear combination of nonlinear dynamics and finishes training quickly by optimizing only the linear combination. 

The performance of RC is evaluated by the MSE or the squared error normalized by the target output. 
The latter is called the IPC~\cite{dambre2012information, kubota2021unifying}. 
To know the approximation accuracy of the relationship between input and output time series, we should evaluate the IPC for infinitely long data, not the IPC for finite-length data. 
Simply evaluating the IPC with long data cannot remove the difference from the limit value. 
We proposed a method to estimate the IPC for infinitely long data by using the asymptotic expansions of the training and the test IPC and the weighted least-squares method. 
We also explained that to determine whether the IPC is $0$, the asymptotic expansion of the variance of the IPC is a robust method. 
Then, we showed the validity of our method by numerical simulations. 
Although we used ESNs in the demonstrations, our method can be applied to RC systems that satisfy the conditions mentioned in the last paragraph of Section 3. 
Since the numerator of the IPC contains the MSE, the proposed method can be applied to estimate the MSE the infinite time limit. 

Our method can estimate the IPC more accurately than simply using long data. 
The drawback of the proposed method is that it is computationally intensive due to multiple trials on long data. 
If we naively implement the proposed method, the computation complexity is approximately $10^3$ times $O(N T d_1^3)$. 
Although we have explained a strategy for reducing the amount of calculation in subsection 3.2, we cannot expect a drastic reduction in computation time, and therefore, the proposed method is considered applicable up to about $d_1 \sim 10^3$. 


We consider a more practical online estimation method which uses Eq.~(\ref{eq:asymp_error_train_IPC}) without taking the sample means. 
Since it is computationally expensive to perform inverse matrix calculations for each $T$ to find $w_T$, we substitute it with $\wt{w}_T$ obtained by the gradient method with a stochastic approximation~\cite{robbins1951stochastic, yuan2024optoelectronic}. 
In Eq.~(\ref{eq:asymp_error_train_IPC}), $\bar{l}_T(w), \bar{\mu}_T$, and $\xi_T$ follow a normal distribution for a sufficiently large $T$, and $\wt{w}_T$ converges to the true parameter $w_0$. 
Thus, if we can estimate the mean of the normal distributions using these terms at different $T$, we can also estimate the true IPC. 
We will leave these for future work. 

\section*{Acknowledgement}
This paper is based on results obtained from a project, JPNP16007, commissioned by the New Energy and Industrial
Technology Development Organization (NEDO). 
We appreciate Professor Takashi Morie, Doctor Tomoyuki Kubota, and Doctor So Nakashima for their fruitful comments. 

\appendix

\section{Weighted least squares method}

First, we derive the asymptotic form of the sample mean of the training IPC. 
Suppose that the number of samples $N$ is large enough. 
Applying the CLT to the sample mean of each term in Eq.~(\ref{eq:asymp_train_IPC}), we obtain 
\begin{align}
 f_1(T) =& 1 - \frac{l(w_0)}{\mu_0} + \frac{1}{T} \left[ \frac{C_{l, \mu}}{\mu_0^2} - \frac{l(w_0) \, V_{\mu, \infty}}{\mu_0^3} + \frac{{\rm Tr}(I_\infty J^{-1})}{\mu_0} \right] 
 + \epsilon_T + o_p\left( \frac{1}{T} \right) \nonumber \\
 =& a + \frac{b_1}{T} + \epsilon_T + o_p\left( \frac{1}{T} \right), 
 \label{eq:asymp_sample_mean_IPC_train} \\
 \epsilon_T \sim& {\cal N}\left( 0, \frac{1}{N T} \left[ \frac{V_{l, \infty}}{\mu_0^2} + \frac{l(w_0)^2 \, V_{\mu, \infty}}{\mu_0^4} - \frac{2}{\mu_0^3} \, l(w_0) \, C_{l, \mu} \right] \right). 
\end{align}
Similarly, the asymptotic form of the sample mean of the test IPC can be written as 
\begin{align}
 f_2(T) =& 1 - \frac{l(w_0)}{\mu_0} + \frac{1}{T} \left[ \frac{T}{T'} \, \frac{C_{l, \mu}} {\mu_0^2} - \frac{T}{T'} \, \frac{l(w_0) \, V_{\mu, \infty}}{\mu_0^3} - \frac{{\rm Tr}(I_\infty J^{-1})}{\mu_0} \right] 
 + \epsilon'_{T'} + o_p\left( \frac{1}{T} \right) \nonumber \\
 =& a + \frac{b_2}{T} + \epsilon'_{T'} + o_p\left( \frac{1}{T} \right),
 \label{eq:asymp_sample_mean_IPC_test} \\
 \epsilon'_{T'} \sim& {\cal N}\left( 0, \frac{1}{N T} \left[ \frac{T}{T'} \, \frac{V_{l, \infty}}{\mu_0^2} + \frac{T}{T'} \, \frac{l(w_0)^2 \, V_{\mu, \infty}}{\mu_0^4} - \frac{T}{T'} \, \frac{2}{\mu_0^3} \, l(w_0) \, C_{l, \mu} \right] \right). 
\end{align}

Next, we obtain $a, b_1$, and $b_2$ using the weighted least squares method. 
Let $\{(T_{i,1}, g_{i,1})\}_{i=1}^N$ and $\{(T_{i,2}, g_{i,2})\}_{i=1}^N$ be the mean IPCs for the training and the test data for each data length. 
We assume that all data lengths are sufficiently long so that the $o_p(1/T)$ terms in Eqs.~(\ref{eq:asymp_sample_mean_IPC_train}) and (\ref{eq:asymp_sample_mean_IPC_test}) can be neglected. 
The least squares method is the maximum likelihood estimation when a normal distribution is used for the likelihood function. 
Since the variance of $\epsilon_T$ and $\epsilon'_{T'}$ is proportional to $1/T$, we use the following cost function, which cancels the change in variance due to the data length; 
\begin{align}
 L(a, b, b') = \frac{1}{2} \sum_{i=1}^N \left[ T_{i, 1} \left(a + \frac{b_1}{T_{i, 1}} - g_{i,1}\right)^2 + T_{i,2} \left(a - \frac{b_2}{T_{i, 2}} - g_{i,2}\right)^2\right]. 
 \label{eq:WLS_cost}
\end{align}
Eq.~(\ref{eq:WLS_cost}) can be interpreted as a cost function that places more emphasis on fitting to the data with less fluctuation. 
Using the notation, 
\begin{align}
 & \beta_1 = \sum_{i=1}^N \frac{1}{T_{i,1}}, \ \ 
 \beta_2 = \sum_{i=1}^N \frac{1}{T_{i,2}}, \ \ 
 s_1 = \sum_{i=1}^N T_{i,1} g_{i,1}, \ \ 
 s_2 = \sum_{i=1}^N T_{i,2} g_{i,2}, \\ 
 & t_1 = \sum_{i=1}^N g_{i,1}, \ \ 
 t_2 = \sum_{i=1}^N g_{i,2}, \ \ 
 \gamma = \sum_{i=1}^N \left( T_{i,1} + T_{i,2} \right), 
\end{align}
we can obtain $a, b_1, b_2$ from 
\begin{align}
 \begin{pmatrix}
  \gamma & N & - N \\
  N & \beta_1 & 0 \\
  N & 0 & - \beta_2 \\
 \end{pmatrix} \, 
 \begin{pmatrix}
  a \\
  b_1 \\
  b_2 \\
 \end{pmatrix} = 
 \begin{pmatrix}
  s_1 + s_2 \\
  t_1 \\
  t_2 \\
 \end{pmatrix} .
\end{align}

Next, we evaluate the variances of the training IPC and the test IPC. 
First, we estimate the fluctuation in the unbiased variance of the training IPC. 
Applying the CLT to Eq.~(\ref{eq:asymp_error_train_IPC}), we find that 
\begin{align}
 & \sqrt{T} (C_T(U'_T) - \mathbb{E}[C_T(U'_T)]) \xrightarrow{d} {\cal N}(0, \alpha), \\
 & \alpha = \frac{V_{l, \infty}}{\mu_0^2} + \frac{l(w_0)^2 V_{\mu, \infty}}{\mu_0^4} - \frac{2}{\mu_0^3} l(w_0) C_{l, \mu}. 
\end{align}
The sample mean of $(T / \alpha) \, (C_T(U'_T) - \mathbb{E}[C_T(U'_T)])^2$ follows $\chi^2(N)$. 
Thus, the distribution of the unbiased variance of $C_T$ can be approximated by the normal distribution whose variance is $(2 N \alpha^2) / ((N - 1)^2 T^2)$. 

Let $f'_1(T) = \frac{d}{T}$ and $f'_2(T) = \frac{d}{T}$ be the asymptotic IPC variances for the training and the test data, respectively, and $\{(T_{i,1}, \sigma^2_{i,1})\}_{i=1}^N$ and $\{(T_{i,2}, \sigma^2_{i,2})\}_{i=1}^N$ be the IPC variances for each data length. 
We employ the cost function with the weight of the variance, 
\begin{align}
 L'(d) = \frac{1}{2} \sum_{i=1}^N \left[ T_{i,1}^2 \left(\frac{d}{T_{i,1}} - \sigma^2_{i,1}\right)^2 + T_{i,2}^2 \left(\frac{d}{T_{i,2}} - \sigma^2_{i,2}\right)^2\right], 
\end{align}
and obtain 
\begin{align}
 d = \frac{\sum_{i=1}^N (T_{i,1} \sigma^2_{i,1} + T_{i,2} \sigma^2_{i,2})}{2 N}. 
\end{align}

\bibliographystyle{elsarticle-num}
\bibliography{biblio}

\end{document}